# Liquid Metal Printed Superconducting Circuits

Wendi Bao [1,2], Jie Zhang [1,2], Wei Rao *[1,2], Jing Liu *[1,2]

1. Key Laboratory of Cryogenic Science and Technology, Technical Institute of Physics and Chemistry, Chinese Academy of Sciences, Beijing 100190, China

2. School of Future Technology, University of Chinese Academy of Sciences, Beijing 100049, China

* E-mail: weirao@mail.ipc.ac.cn; jliu@mail.ipac.cn

## Abstract

Since the discovery of superconductor one hundred years ago, tremendous theoretical and technological progesses have been achieved. The zero resistance and complete diamagnetism of superconducting materials promise many possibilities in diverse fields. However, the complexity and expensive manufacturing costs associated with the time-consuming superconductor fabrication process may retard their practices in a large extent. Here, via liquid metal printing we proposed to quickly fabricate superconducting electronics which can work at the prescribed cryogenic temperatures. By way of the room temperature fluidity of liquid metal composite inks, such one-step printing allows to pattern various superconducting circuits on the desired substrate. As the first-ever conceptual trial, the most easily available gallium-based liquid alloy inks were particularly adopted to composite with copper particles to achieve superconductivity under specific temperatures around 6.4K. Further, a series of liquid metal alloy and particles loaded composites were screened out and comparatively interpreted regarding their superconducting properties and potential values as printable inks in fabricating superconducting devices. The cost-effective feature and straightforward adaptability of the fabrication principle were evaluated. This work suggests an easy-going way for fabricating ending user superconducting devices, which may warrant more promising investigations and practices in the coming time.

# 1. Introduction

Superconductors exhibit many intriguing properties such as zero resistance, absence of heating, and high current transport with no loss. These features make them promising for the development of high-speed micro/nanoscale electronic devices, including fast-response biosensors, quantum computers, and micronuclear magnetic resonance devices. The earliest superconductors were discovered in liquid metal mercury (Hg), which however would present its environmental toxicity and thus limit the practical value [1]. Since then, investigations on superconductors have been mainly focused on solid superconducting materials, such as alloy superconductors, copper-based superconductors, and iron-based superconductors[2–5], with a lack of reports on superconductivity in other liquid metals. However, conventional complex preparation and processing treatments of superconductors were often in-efficient. Despite some recent progress in fabricating superconducting electronic devices by photolithography[6,7], their costs are relatively high.

In addition, flexible applications of superconducting devices, including superconducting tapes and wires, superconducting thin films, and some micro-superconducting devices, have received increasing attention over the years. In the field of superconducting thin films like sputtering, electron beam evaporation, pulsed laser deposition, and chemical vapor deposition, high vacuum equipment is required, and the film deposition process is very time-consuming. Recently, the commercial application of superconducting thin films using the coating process has been widely investigated, and large-area films can be fabricated easily and quickly without needing high vacuum equipment[8]. Whereas, the preparation of most superconducting thin film materials is fabricated on rigid substrates, which limits the flexible applications of superconducting materials and devices. So far, the development of superconductors that can be applied to microelectronic devices with comprehensive properties such as high critical temperature, good flexibility, and easy processing is still a big challenge, which is also an inevitable requirement for the wide uses of flexible superconducting microelectronics.

With the emergence of room-temperature liquid metal science and technology[9], many new properties within such materials were revealed. Importantly, the low-temperature superconductivity in gallium-based alloys also received the attention of researchers[10]. Gallium-based alloys are low-melting-point metals that are room-temperature fluid, non-toxic, easy to process and fabricate, and capable of rapid patterning, allowing for easy and fast fabrication of superconducting films and circuits, making them suitable for applications in superconducting micro-flexible devices. The critical temperature $T_c$ of gallium-based alloys is above the liquid helium temperature (4.2 K, the lowest temperature for practical application), which makes this flexible material potentially useful in superconducting electronic fields. Liquid metal printing provides straightforward ways for the rapid preparation of functional circuits using the direct fabrication strategy[11,12], by utilizing the adhesion property of liquid metal which significantly innovating the realization of microelectronic devices. Figure 1 summarizes some typical liquid metal inks and their flexible superconductivities. The method of doping particles has become a mainstream manufacturing method for liquid metal printing ink due to its simple preparation and outstanding material properties. Meanwhile, flexible superconducting materials, including superconducting nanowires[10,13–15] and superconducting films on the flexible substrate[16–19], have also been developed. Furthermore, scientists have embedded superconducting wires into elastic substrates and successfully designed flexible superconducting PCB circuit boards[20], as well as superconducting resonators[21–23]. So far, introducing liquid metal inks for directly printing superconducting devices has not been tried.

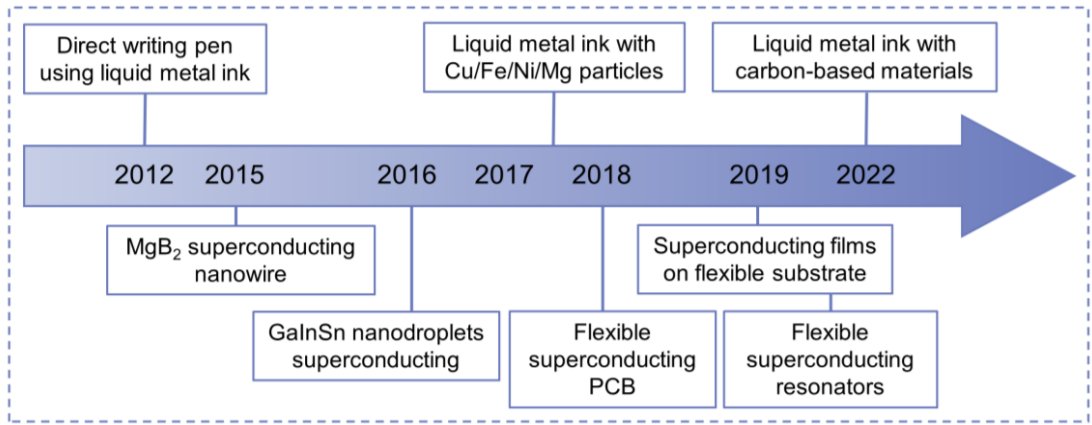

**Figure 1** Timeline of main typical works on liquid metal inks and flexible superconducting items.

In this work, we proposed to directly fabricate superconducting electronic devices by liquid metal printing. Through doping Cu particles in liquid metal EGaInSn, liquid metal composite ink (Cu-EGaInSn) as prepared can adhere well to the target substrate and construct the desired superconducting circuit patterns. The superconducting properties of the liquid metal film were then evaluated and mechanisms were interpreted. The inks can contribute to the rapid fabrication of superconducting electronic patterns and circuits, as well as realization of superconductivity at specific cryogenic temperatures. Such endeavor suggests an efficient method for fabricating end-user superconducting devices and provides a promising avenue for combining superconducting materials with flexible devices.

## 2. Materials and Methods
### 2.1 Materials

In this work, the used GaInSn liquid metal is EGaInSn composed of 67 wt% gallium, 21.5 wt% indium and 12.5 wt% tin (Anhui Minor New Materials Co. LTD, 99.999% purity). The copper microparticles with a diameter of 20 μm were purchased from Beijing DK Nano Technology Co., Ltd. PDMS film substrates were purchased from Juancheng Electronic Accessories Co., Ltd.

### 2.2 Preparation of Cu-EGaInSn

The preparation process is shown in Figure 2a. At first, 20g of EGaInSn was placed in a beaker. Then, 4g of Cu microparticles was added to the surface of EGaInSn contained in the beaker. Later, the NaOH solution (1.0 mol L$^{-1}$) in volume to EGaInSn was transferred to the beaker and stirred for 1 min via a glass bar. Letting the sample stand still for 30 min to help the copper microparticles be completely internalized into the EGaInSn. Finally, the NaOH solution was evaporated at 70 °C for 2h in the vacuum drying oven. The EGaInSn mixed with Cu particles was thus obtained (Figure 2b) and characterized (Figure 2c), named Cu-EGaInSn ($\Phi = m_{Cu}/m_{EGaInSn} = 20\%$, where $m_{Cu}$ and $m_{EGaInSn}$ represent the mass of Cu particles and EGaInSn, respectively).

### 2.3 Liquid metal printing of superconducting circuits

Fabrication of the Cu-EGaInSn superconducting circuits first required designing the circuit and engraving it onto A4 paper using laser as a paper mask. Then the mask was placed on the PDMS substrate and Cu-EGaInSn was evenly printed onto the substrate by a brush dipped in the liquid metal. After removing the mask, the desired circuit was obtained (Figure 2d). The circuit could deform and stretch at room temperature (Figure 2e), and showed superconducting properties below the superconducting transition temperature.

### 2.4 Superconducting measuments

To characterize the superconducting properties of the liquid metal printed circuits spanning wide range of temperatures, a physical property measurement system (PPMS) was adopted. For the resistance measurement, patterned samples were printed using a mask method on a non-conductive PDMS substrate as 2mm * 10mm lines with a thickness of approximately 0.1mm. The standard four-wire method was used to measure the sample's DC resistivity, which requires silver wires welded to a dedicated sample holder, as shown in Figure 2f. For magnetization measurements, the vibration sample magnetometer (VSM) of PPMS was adopted. It should be pointed out that, the sample needs to be pre-frozen for placement into the VSM chamber.

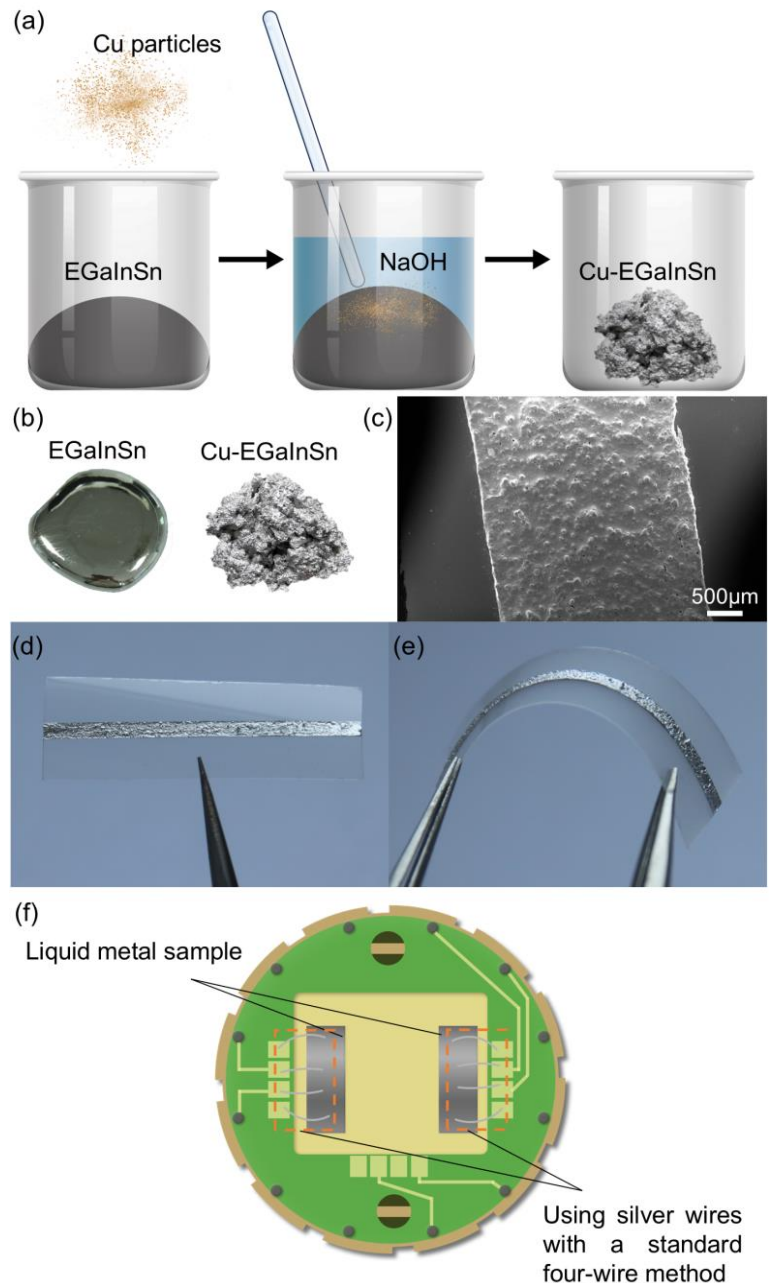

**Figure 2** The preparation and characterization of the Cu-EGaInSn. (a) The preparation process of Cu-EGaInSn. (b) The optical images of EGaInSn and Cu-EGaInSn. (c) Microscopic characteristics of Cu-EGaInSn printed on PDMS substrate. (d) The Cu-EGaInSn line was printed on a PDMS substrate through a direct printing method. (e) The Cu-EGaInSn line under bending. (f) Schematic diagram of a sample holder for the resistance measurement of PPMS.

## 3. Results and Discussion

### 3.1 Characterization of Cu-EGaInSn

The optical images of EGaInSn and Cu-EGaInSn in Figure 2b indicate a significant decrease in the fluidity of Cu-EGaInSn. As shown in Figures 2d, 2e, Cu-EGaInSn displays high adhesion on the PDMS substrate, and the printed circuit can maintain a stable electrical connection and good conductivity after multiple bending and stretching.

Furthermore, the microscopic characteristics of Cu-EGaInSn printed on a PDMS substrate were studied (Figure 2c). At present, the thickness of Cu-EGaInSn on PDMS is about 0.2mm. From the image, some $CuGa_2$ particles in EGaInSn make the surface of Cu-EGaInSn appear uneven.

Further investigation of its composition was conducted using an X-ray diffractometer (XRD), as shown in Figure 3a. The XRD peaks showed the presence of $CuGa_2$ in the Cu-EGaInSn sample, indicating that during the doping process, Cu reacted with EGaInSn to form $CuGa_2$ particles, which is also the reason for the rough surface of the printed sample.

The thermal properties of Cu-EGaInSn were tested using differential scanning calorimetry (DSC). In Figure 3b, the freezing point of Cu-EGaInSn is 257K, lower than its 293K melting point. The difference between the two points is caused by the supercooling characteristics of the liquid metal. Supercooling is the process that liquid can be cooled below its freezing point instead of becoming solid. The supercooling of the Cu-EGaInSn sample is 36K. Also, the freezing point of Cu-EGaInSn is higher than that of the pure EGaInSn (Figure 3c), while the melting point of Cu-EGaInSn is lower than that of the pure sample. The supercooling value of the EGaInSn can even reach 80K. It is speculated that the introduction of Cu inhibits the supercooling of the liquid metal because of the addition of nucleating agents[24,25].

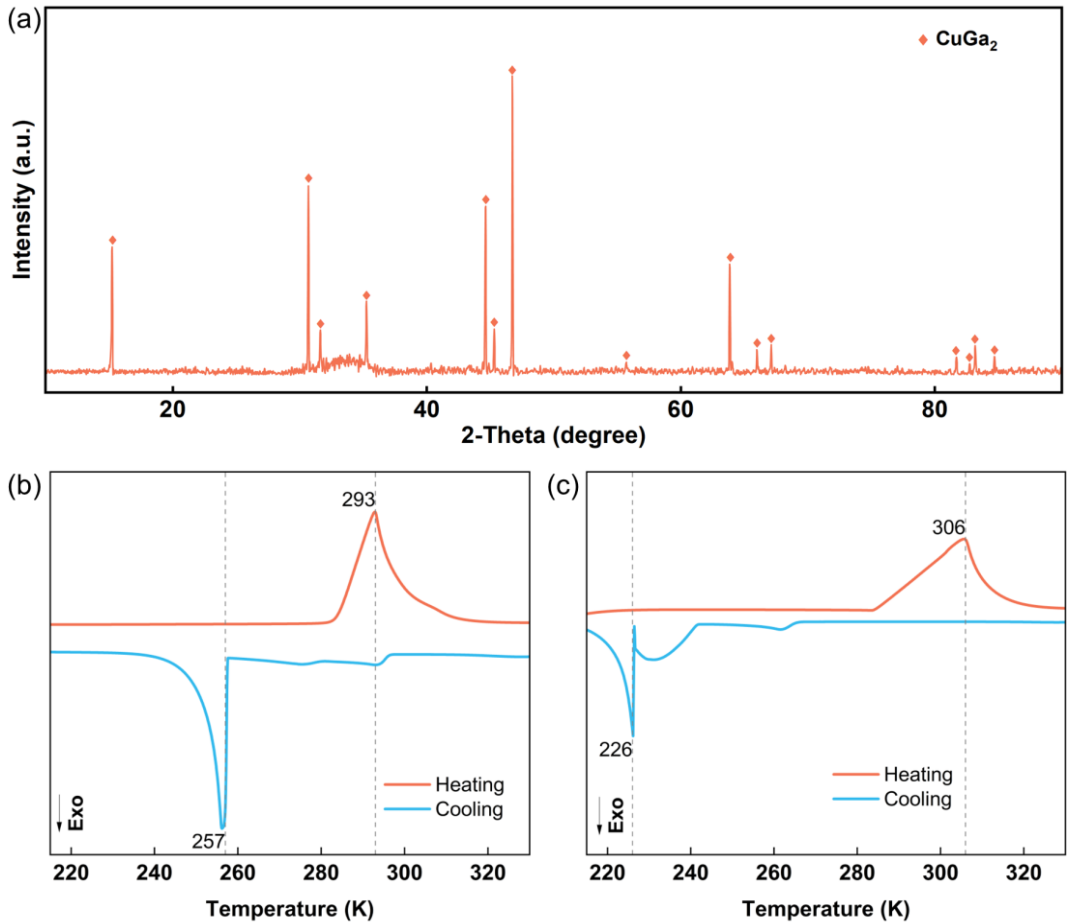

**Figure 3** Characterization of Cu-EGaInSn. (a) XRD patterns for the Cu-EGaInSn at room temperature. The rhombic symbols indicate the Bragg reflections due to the $CuGa_2$. (b) The DSC curves for the Cu-EGaInSn. The red line indicates the heating curve, while the blue line depicts the cooling curve. (c) The DSC curves for the pure EGaInSn. The red line shows the heating curve, while the blue line displays the cooling curve.

## 3.2 Superconductivity of Cu-EGaInSn

This study revealed the superconducting performance of Cu-EGaInSn at low temperatures. Figure 4a shows the temperature-dependent (R-T) curve of the resistance for Cu-EGaInSn printed circuits. It can be observed that, the sample exhibits a resistance transition at 6.4K, reaching a superconducting state. This is higher than the superconducting transition temperatures of the individual components (gallium at 1.08K, indium at 3.41K, tin at 3.72K). According to the GaInSn ternary phase diagram, a conventional eutectic reaction occurs in the proposed EGaInSn alloys, the $In_3Sn$ phase ($T_c$ at 6.6K) and $InSn_4$ phase ($T_c$ at 4.4K) form below 260K[26–28]. This indicates that the main superconducting phase formed in the sample is $In_3Sn$, and $InSn_4$ with the lower superconducting temperature as a secondary phase. Notably, there were jumps in the R-T curve of the sample in the cooling and heating processes, which may be caused by the phase separation of the sample. During the cooling process, due to the transition of the amorphous matrix to a crystallized state in the system, ordered phonon vibrations are formed, and the scattering of charge carriers is weakened, increasing conductivity, which is manifested as a sudden decrease in resistance as the temperature decreases. When cooling and heating, the resistance changes sharply around 257K and 293K, respectively. This phenomenon is consistent with the solidifying and melting behavior observed in DSC testing. In addition, we also observed a jump in resistance around 100K during the cooling process, similar to the jagged behavior observed by Mochiku et al[27], which may be attributed to the mutual influence between the base material and EGaInSn material.

Figure 4b shows the magnetization temperature curve of Cu-EGaInSn sample under a 50Oe magnetic field. The separation of the zero-field cooling (ZFC) curve and the field cooling (FC) curve occurs at a superconducting transition temperature $T_c$ of 6.4K, which matches the R-T curve. Temperature dependence of the resistance and magnetization curves both prove that the Cu-EGaInSn sample prepared in this work improves its printing ability and keeps its superconducting performance.

Figure 4c displays a comparative study of the temperature dependence of the resistance between EGaInSn and Cu-EGaInSn. Comparing the copper-doped samples with the pure samples, both exhibit similar behaviors between 2K-300K and almost the same superconducting temperature, which proves that the introduction of copper can have a nearly negligible effect on the superconducting properties of the liquid metal while increasing their printing properties. Figure 4d illustrates the superconducting temperatures of several typical liquid metals. On the whole, the superconducting temperature generally displays a progressive trend from single element to binary alloy, and then to ternary alloy. Hg, first discovered to be the superconductor, has rarely been studied due to its toxicity. Besides, pure Ga, In, Sn have little application value as superconductor due to their superconducting temperature being lower than that of liquid helium. However, with pretty high melting points, the processing and application of binary alloys $In_3Sn$ and $InSn_4$ are limited compared to room-temperature liquid metals. The superconducting temperature of EGaInSn is affected by the formation of binary phases at low temperatures, which is the same as the superconducting temperature of binary alloys. The focused Cu-EGaInSn samples in this work have similar Tc values with In₃Sn while being capable of direct printing, which warantees potential application value in the field of superconducting electronics in the near future.

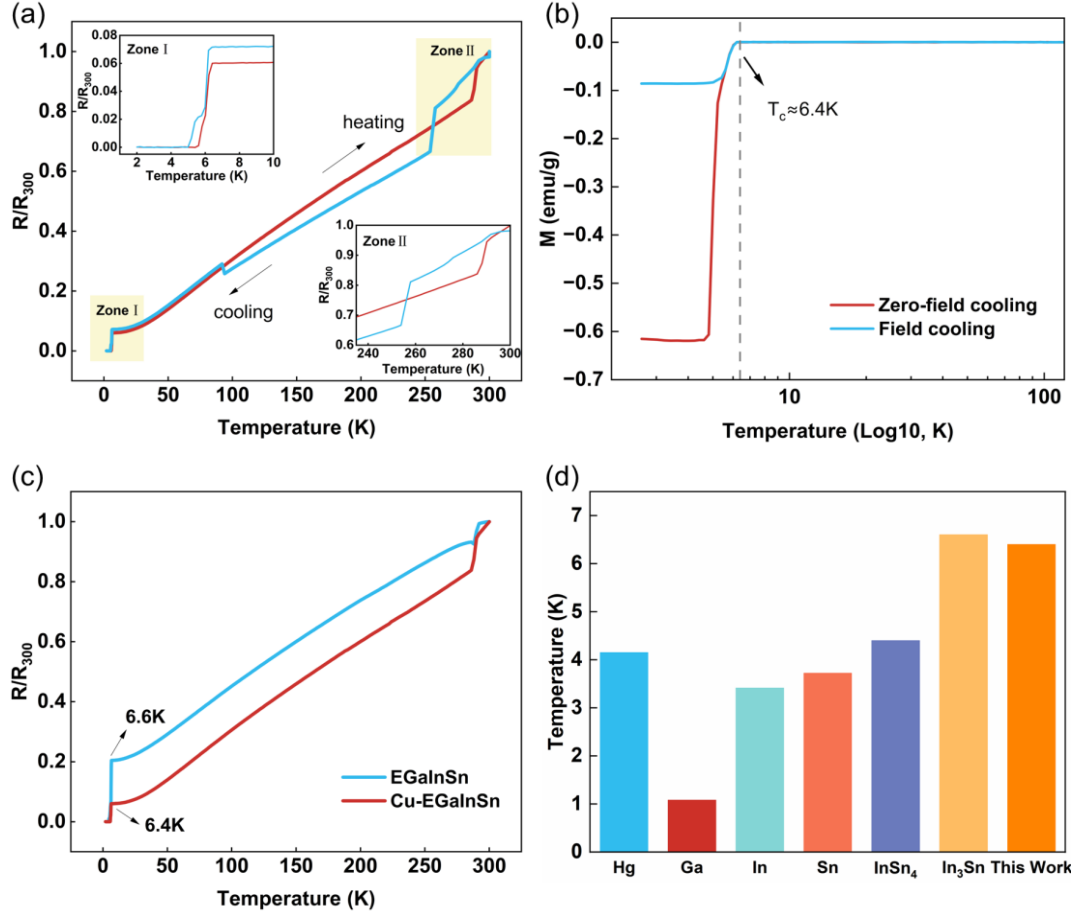

**Figure 4** Superconductivity of the liquid metals. (a) Temperature dependence of the resistance (R – T curves) between 2K and 300 K for the printed Cu-EGaInSn ink. The red line represents change in resistance as the temperature increased, while the blue line represents the resistance change as the temperature decreased. The insets are the zoom-in images at the Zone I and Zone II area, showing the superconducting transition and the phase change regions respectively. (b) Temperature dependence of zero-field cooling and field cooling (ZFC, FC) magnetization from 2K to 300 K of the Cu-EGaInSn under a magnetic field of 50 Oe. (c) Temperature dependence of the resistance between 2K and 300 K for the EGaInSn and the Cu-EGaInSn, comparing their superconducting properties. (d) Several physical properties of certain superconductors.

We discuss the effect of alloying on the superconducting system through the BCS formula which reads as:

$$T_c = \alpha \theta_D \exp[-(1/N(0)V)] \quad (1)$$

where $\theta_D$ is the Debye characteristic temperature of the solid, $\alpha$ is a numerical constant of order unity, $N(0)$ is the density of states in energy at the Fermi energy, and $V$ is a measure of the strength of the electron - phonon coupling. In the study of nontransition state alloy systems, it was found that the variation of V value with alloy composition has a certain universality, and once the average free path of electrons becomes less than the coherence range, alloying can always promote an increase in $T_c$[29], This is also consistent with the conclusions we have reached in the previous section. However, the slight decrease in the superconducting temperature of the Cu-doped sample compared to EGaInSn can be attributed to band filling and interband scattering effects[30]. The introduction of electrons and holes has a band filling effect on the energy bands, directly altering the density of states on the Fermi surface and thus reducing $T_c$. While interband scattering of the Cu affects the electron-phonon interaction, which may also lead to a decrease in $T_c$.

Further, some candidates for low melting point metal superconducting materials and printable liquid metal superconducting materials were collected and listed in Table 1 and Table 2, respectivey. They are expected to be used as inks for printing flexible superconducting circuits in subsequent experiments. In the coming time, during the search for new candidate materials, traditional liquid metal inks, and new high entropy alloys can both be explored. In addition, to improve the efficiency in discovering future superconducting materials along this direction, screening experiments and evaluations can be combined with simulation calculations based on the constructed material databases and machine learning methods.

Table 1 Certain low melting point metal superconducting materials.

| Element or alloy | $T_m$ (°C) | $T_c$ (K) | $\sigma$ (×10$^6$ S m$^{-1}$) | Toxic | Ref. |
|---|---|---|---|---|---|
| GaIn$_{20.5}$Sn$_{12.5}$ | 10.5 | 6.6 | 3.1 | No | This work |
| GaIn$_{15}$Sn$_7$ | - | 6.0 | - | No | [27] |
| GaIn$_{21.5}$Sn$_{10}$ | ≈10.0 | - | 3.3-3.5 | No | [31] |
| GaInSn (62:25:13 by weight) | - | 6.28 | - | No | [32] |
| EGaIn | 15.5 | - | 3.4 | No | [31] |
| InGa$_{0.04-0.12}$ | 120-140 | ≈3.4 | - | No | [33,34] |
| InSn$_{0.02-0.98}$ | 117-230 | 3.5-6.6 | - | No | [35,28] |

| Element or alloy | $T_m$ | $T_c$ (K) | σ (×10⁶ S m⁻¹) | Superconducting | Ref. |
|---|---|---|---|---|---|
| SnGa$_{1-5}$ | 220-230 | 3.7-3.9 | - | No | [36,37] |
| GaSn$_{13.8}$ | ≈20.5 | - | 3.3 | No | [31] |
| GaZn$_{3.6}$ | ≈25.2 | - | - | No | [31] |
| GaSn$_{10.8}$Zn$_{2.9}$ | ≈8.0 | - | 2.95 | No | [31] |
| GaIn$_{25}$Sn$_{13}$Zn$_1$ | ≈7.6 | | | No | [31] |
| BiIn$_{51}$Sn$_{16.5}$ | ≈60.0 | - | - | No | [31] |
| Pure Ga | 29.76 | 1.08 (α-phase) / 6.2 (β-phase) | 3.7 | No | [38,39] |
| Pure In | 156.6 | 3.41 | 12.5 | No | [31] |
| Pure Sn | 231.9 | 3.72 | 8.7 | No | [31] |
| Pure Hg | -38.37 | 4.15 | 1.04 | Yes | [40,41] |

Note: $T_m$: melting point; $T_c$: Superconducting transition temperature; σ: Conductivity

Table 2 Candidates of printable liquid metal superconducting materials.

| Element or alloy | Fraction of additive | $T_c$ (K) | σ (×10⁶ S m⁻¹) | Ref. |
|---|---|---|---|---|
| Cu-EGaInSn | 20 wt.% | 6.4 | 3.5 | This work |
| Zn-GaInSn | 1.7 wt.% | 6.06 | - | [32] |
| Fe-EGaIn | 40 vol% | - | 3.9 | [42] |
| Ni-EGaIn | 15 wt.% | - | 1.4 | [43] |

## 4. Conclusion

In this work, we had proposed a straightforward and convenient method for rapidly manufacturing superconducting electronic circuits. It was demonstrated that the liquid metal composite ink (Cu-EGaInSn) prepared by doping functional powders with a liquid matrix can be successfully printed to an elastic substrate, making it a stretchable flexible circuit at room temperature and achieving superconductivity below 6.4K. This principle has universality and can develop promising superconducting devices based on liquid metal solvents and more printing technologies in the near future. It is expected to serve as a fast and low-cost method for manufacturing micro superconducting devices.